\documentclass[12pt,a4paper]{article}
\title{\bf Third Harmonics and Defects \\ in Modulated Phases of Structural \\ Quantum
Order-Disorder Systems.}

\author{Matej Hudak \\ {\it Matej Hudak's Lab A,} Stierova 23, SK - 040 23  Kosice, Slovak Republic  \\hudakm@mail.pvt.sk
\and Jana Tothova\\ {\it Matej Hudak's Lab A,} Stierova 23, SK - 040 23  Kosice, Slovak Republic
\and Tatiana Hudakova\\ {\it Matej Hudak's Lab A,} Stierova 23, SK - 040 23  Kosice, Slovak Republic
\and Ondrej Hudak \\  {\it Matej Hudak's Lab A,} Stierova 23, SK - 040 23  Kosice, Slovak Republic\footnote{Corresponding author}\\
{hudako@mail.pvt.sk}}
\date{May 26, 2019}
\begin{document}
\maketitle{}

\section*{Condensed paper title:}

{\it Third Harmonics - Defects in Modulated Phases - Structural Quantum
Order-Disorder Systems.}
\newpage
\vspace{1in}
\begin{abstract}
Influences of defects generating random fields on modulated phases in materials with the order-disorder type of phase transition are studied. It is shown that under appropriate conditions third harmonics of the ground state modulation may be suppressed by these defects.
\end{abstract}
\newpage

\tableofcontents{}

\newpage
\section{Introduction.}

Incommensurate phases in ferroelectrics attracted interest of physicist due to many interesting properties \cite{tfc}. As concerning theoretical and experimental methods \cite{WBJW} Handbook of Physics is a complete reference for scientists, engineers, and students for everyday use. Incommensurate-commensurate states phenomena are interesting phenomena, for example in these days they are concerning topological quantum matter \cite{AMEC} in studying commensurate and incommensurate states of topological quantum matter. Defects in general play important role in many systems. For example in \cite{YLMAG} Lansac, Glaser and Clark study discrete elastic model for two-dimensional melting. They study both geometrical defects (localized, large-amplitude deviations from ordering) and topological defects (dislocations and disclinations), as they play a role in 2D melting. Their model exhibits a rich phase behavior including hexagonal and square crystals, expanded crystal, dodecagonal quasicrystal, and liquid structure. 
In \cite{GDGHR} ordered and disordered defect chaos is studied
numerically in coupled Ginzburg-Landau equations for parametrically driven waves. The motion of the defects is traced in detail yielding their life-times, annihilation partners, and distances traveled. Rubtsov and Janssen \cite{ANRTJ} study thermodynamic properties of the discommensuration point for incommensurate structures. The consequences of the opening of a phason gap in incommensurate systems are studied on a simple model, the discrete frustrated $\phi^{4}$-model.

In general together with the first harmonic response there may be higher order harmonics in the response to external fields. For exmple in ferroelectrics there exist glass states. Third and fifth harmonic responses are reviewed in \cite{SAMM} glasses, namely $\chi_{3}$ and $\chi_{5}$. Authors explain why these nonlinear responses are especially well adapted to test whether or not some amorphous correlations develop upon cooling. They show that the experimental frequency and temperature dependences of $\chi_{3}$ and of $\chi_{5}$ have anomalous features, since their behavior is qualitatively different to that of an ideal state. Most of the works have interpreted this anomalous behavior as reflecting the growth, upon cooling, of amorphously ordered domains, as predicted by the general framework. Finally, the comparison of the anomalous features of $\chi_{5}$ and of $\chi_{3}$ shows that the amorphously ordered domains are compact, i.e., the fractal dimension $d_{f}$ is close to the dimension $d$ of space. This suggests that the glass transition for example of molecular liquids corresponds to a new universality class of critical phenomena. The same type of study may be realized for glass states in ferroelectrics. That this study would be useful for ferroelectrics where highly-efficient second and third harmonic generation in a monocrystalline lithium niobate microresonator is used \cite{JLYN}, and where it would be interesting to study how defects this generation change.

In this type of materials and its solid solutions \cite{AYBPKD}
 the structure and dielectric properties of \\ the  $Li_{1+x-y}Nb_{1-x-3y}T_{x+4y}O_{3}$
system have been investigated. Detailed studies of the lattice
parameters of these phases agree well with structure models based on intergrowths of $LiNbO_{3}$ slabs with a Ti-rich corundum-type layer. The relative permittivity ranges from $-80$ to $-55$. The temperature coefficient of the resonant frequency, $τ_{f}$ , changes sign within the solid solution region thus permitting tunability to a zero value. $LiNbO_{3}$ is interesting because of its
ferroelectric and opto-electronic properties. Neither of the previous studies was able to refine the structures in detail, but two models were proposed. In one model,
developed by Smith and West, the solid solutions were described as incommensurate structures comprised of random intergrowths of $LiNbO_{3}$-type and rocksalt-type $Li_{2}TiO_{3}$ slabs. This model also predicted that random stacking faults in the anion packing are associated with the interfaces between the two blocks. In a second model Roth et al. proposed the phase field is comprised of a homologous series of commensurate structures, based on ordered intergrowths of $LiNbO_{3}$ and a spinel-type end-member $Li_{4}Ti_{5}O_{12}$. As we can see there exist incommensurate phases in which defects may occur. It would be interesting to know whether measurements of the third order susceptibilities are possible to measure.

Defects presence
leads to such effects like global hysteresis \cite{31} and memory effects
\cite{32}. Order-disorder systems, like dipolar glasses of the $RADP$ and
\( Rb_{1-x}(NH_{4})_{x}H_{2}PO_{4} \) type contain defects conveniently
described by the random fields and random interactions \cite{33}. Spin models in weak random fields are good representations of a large number of impure compounds which undergo magnetic or structural transitions. From a theoretical point of view these systems exhibit frustration on a mesoscopic scale and take therefore an indermediate position between (spin) glasses and unfrustrated (diluted) disordered systems.

There are two types of frustration in the dipolar glasses: the frustration
due to competing between ferroelectric-antiferroelectric structural ordering and
the frustration due to competing between different local random fields acting
on the collective ordering of the fluctuating quantum units (hydrogen bonds,
deuteron bonds, ...).
The phase diagram of the mixed crystals $RADP$ and $DRADP$ showing ferroelectric, antiferroelectric and
dipolar glass phases is the best example of real physical system with
mentioned two types of frustration. There are randomly distributed ferro- and antiferro-
electric local interactions due to distribution of \( Rb^{+} \) and
\( NH_{4}^{+} \)
cations. Two different orderings are associated with the two types of cations.
In the glass region this type of competition would lead to long range modulated
ordering, \cite{5}. In real systems short-range correlations freeze only.
A model of the hydrogen (deuteron) bonding \cite{3} was developed, it is
able to describe the incommensurate wavevector behavior as well as some other
properties of the glass phase. Far-infrared and near-millimetre spectroscopy,
\cite{10} has shown that above the glass transition temperature the
low frequency dynamics is similar to that of an averaged pure crystal.
Below this temperature a broad background absorption in addition to phonon peaks
is present. The appearance of new forbidden modes below the glass transition
temperature gives, according to the authors, evidence of polar microregions.
These modes \( B_{2} \) are forbidden in the paraelectric phase but, at least
below the glass transition temperature, they are observed.
Another example is the influence of partial bromination on the
$(p,T)$-diagrams. Dielectric and optical measurements, \cite{MLM 92},
have shown that an increase of the $Br$ concentration leads to decrease
and finally to the disappearance of the stability regions of more and
more longperiod phases in $BCCD$. It would be interesting to study
qantitative changes in modulation of the dissappearing structures.

A method for measuring the nonlinear response in dielectric spectroscopy through third harmonic detection is described in \cite{ctdlh}. This is a high sensitivity method allowing the measurement of the nonlinear dielectric susceptibility of an insulating material at finite frequency.

To interpret these and other observations, one should understand better
influence of defects on modulated phases in materials with
the order-disorder type. This understanding should be based on the
microscopic level detailed models of c
qualitative considerations based on simplified models are sufficient
to discuss existing experimental evidences for  $\chi_{3}$ and of $\chi_{5}$  without defects.

The ideal systems without defects are described by the quantum Ising-like Hamiltonian in the simplest cases. The main aim of this paper is to study influence of a few
on-site defects on the ground state properties of incommensurate systems in quantum order-disorder type materials using expansions in a small parameter - the pseudospin
amplitude.

In the next section we describe sinusoidaly modulated incommensurate structure in quantum structural order-disorder type materials as idealized crystals.
Then we introduce the model, which we study. In the fourth section a description of the modulated ground state in the idealized medium without defects is presented for reference purposes. The fifth section contains description of defects and their influence on the groud state properties in modulated phases. We discuss the simplest cases of a single defect and of a pair of defects. The sixth section is devoted to the incommensurate phase properties changed due to presence of defects.
In the last section we discuss our results in connection to a general theory and to related experimental systems.

\section{Sinusoidaly modulated incommensurate structure in quantum structural order-disorder type materials as idealized crystals.}

Below we studied interaction of a single defect and of a pair of defects,
characterized by associated random fields, with ordering tendencies
of fluctuating units described by an Ising type interactions between
nearest neighbors and next nearest neighbors. We neglect tunneling
phenomena in our approach to simplify calculation as much as possible.
Incorporated are two types of frustration: due to competition between
ferroelectric and antiferroelectric interactions and due to local
random fields of defects. For recent review on random field
systems see in \cite{33}.
Our basic questions are: how presence of defects influence ground
state ordering and dynamics above the order-disorder transition temperatures,
how the incommensurate wave developed below this transition temperature
will be modified by defects, is it true that defect main effect is to lock-in
the phase of the wave or there are changes of the wave amplitude and wave vector
in the neighborhood of defects.

Ferroelectrics
and antiferroelectrics, in which
quantum motion of some units is responsible for an order-disorder
type phase transition are described by using pseudospin formalism.
The modulated single plane wave (sinusoidal) ground state
structure may be stabilized below some critical temperature \( T_{c} \),
\cite{14}.
At nonzero wavevector \( {bf q}_{0} \) there exists maximizing
the quantity \( J_{{bf q}} = \sum_{j} J_{ij} \exp( i{bf q}.({bf R}_{i}
-{bf R}_{j}) \) in this case.
Here \( J_{ij} \) are the interaction energies
between two units localized at lattice sites \( {bf R}_{i} \) and
\( {bf R}_{j} \) in the most simple theories.
Dynamics of such a system is described
by the quantum Ising-like Hamiltonian.
The excitation spectrum and response of the later
model was recently studied for example in \cite{15}, \cite{13} and by the authors of
this paper \cite{DHH}.

The freezing phenomena of reorienting moments in solids leading to an
orientational glass phase were studied thoroughly recently, \cite{1}.
Clustering phenomena are observed in some materials, like in
\( Rb_{1-x}(NH_{4})_{x}H_{2}PO_{4} \), which are solid solutions of
hydrogen-bonded ferroelectric and antiferroelectric compounds.
Satellites occurring in X-ray scattering at some specific points of the
Brillouin zone reflect tendency of these systems towards a superstructure,
\cite{2}. This superstructure may be characterized by the wavevector
\( q=0.15a^{*} \) for \( x \geq x_{c}, \) here \( x_{c} \) is a material
constant, characteristic value of which is \( \approx 0.2 \).
With increasing concentration
x of the ammonium group the modulation wavevector q slightly increases,
\cite{3}: for \( x=0.35 \) it is \( q \approx 0.25a^{*} \), for x=0.49 it is \(
q \approx
0.32a^{*} \) and for x=0.68 and x=0.75 it is \( q \approx 0.35a^{*}. \)
At 160K and above this temperature the modulation disappears. Deuterated
isostructural compound, \( Rb_{1-x} (ND_{4})_{x} D_{2}PO_{4}, \) also
shows modulation at a wavevector of approximately \( 0.3a^{*}, \) see \cite{4}.
A pseudospin model appropriate for this later compound is formulated in
\cite{4}.
An incommensurate phase between the ferroelectric (x=0) and the
antiferroelectric
(x=1) phases is predicted to be stable around \( q \approx 0.35a^{*} \)
for a wide range of concentrations of \( ND_{4} \) group. Outside this value
of concentrations the wavevector q of the resulting structure tends to
the uniform modulation q=0 for x=0 and to the antiferroelectric for x=1.
The final glass behavior follows from the presence of random fields due to
random
distribution of lattice sites in which \( ND_{4} \) substituents are present.
Deuterated compounds are satisfactorily described neglecting quantum tunneling
phenomena.
Thermal neutron scattering experiments on this later compounds are
characterized by the presence of the cigar-shaped diffuse peak on
\( \Sigma \) line, \cite{5}. A freezing temperature of about 100K
is found. Below this temperature the spatial correlations stop to
increase. Cigar-shape peaks arise from correlations which are different
in different directions. For x=0.62 these correlations are larger in
the (1,0,0) direction and peaked at about \( q= \frac{1}{3} a^{*} \),
\cite{6}. A broad \( ( \nu \le 0.5 \) THz ) inelastic contribution is
observed at an arbitrary point of the Brillouin zone. The dynamics of the
relaxation may be characterized by two components with different ways
to the frozen state, \cite{7}. The effective medium would condense at an
(incommensurate) wave vector q. However the condensation of short
wave-length fluctuations in regions separated by substituents lead to
the state which is determined by conflicting boundary conditions
due to substitutional disorder. As a consequence the glassy state
results with regions slowing down of which is described by the Vogel-Fulcher
type of dependence, \cite{1}.

The complex dielectric constants in $RADP$ ($x=0.35$) and $DRADP$ ($x=0.35$) have
been described also by the Vogel-Fulcher law of the relaxation time \cite{8}. Considerable dielectric dispersions are observed at low temperatures
for both $RADP$ and $DRADP$ samples, deuteration shifts the dispersion
region to lower frequencies at the same temperature.
The dispersion is
observed in a wide range of frequencies (from 11Hz to 1THz in \cite{8} ).
Such unusual
low temperature and low frequency behavior reflects also
in the thermal properties of the system. Glass-like excitations are
observed in low temperature thermal conductivity \( \kappa \approx T^{3} \)
and in the specific heat \( C \approx T \) in $RADP$ ($x=0.32$, $x=0.72$) below
100K, \cite{9}. In $DRADP$ ($x=0.62$) the same authors did not observe the
glasslike behavior in the thermal conductivity while it was still present
in the specific heat at very long times. Such an isotope effect is
attributed in \cite{9} to the low energy excitations in these crystals related
to tunneling of protons or deuterons. A broad background absorption in addition
to
phonon peaks remain present below the freezing (100K) temperature down to lowest
temperatures in the $RADP$ ($x=0.5$), \cite{10}, which is probably caused
by breaking of the quasi-momentum conservation selection rules
due to incommensurability of this system.
In the isostructural $DRADP$
crystal ($x=0.5$) no background absorption is observed in the temperature
range $15 K-300 K$, \cite{11}. The absence of the background absorption in $DRADP$
in this temperature range may be explained, \cite{11}, by
higher diffusivity of the walls between clusters in this compound than that
in $RADP$. It is noticed in \cite{11} that the mode damping is still high
at such low temperatures at which the anharmonic deuteron (proton) motion
(hopping) is believed to be frozen in. This behavior is in sharp contrast
with that for pure $RDP$ and pure $ADP$ compounds where the damping constants
are characterized by considerably lower values.
In contrast to pure $RDP$ and $ADP$ the soft mode in $RADP$ and $DRADP$ exhibits
continuous softening on cooling down to $5 K$. The soft relaxation
frequency in $DRADP$ is one order lower than in $RADP$, confirming the proton
 (deuteron) origin of this mode.
Dielectric dispersion
measurements \cite{12} do not see any obvious x dependence in the dynamic
properties of $RADP$.
Not all of the mentioned low temperature characteristics are explained
within the existing theories, see in \cite{1}.

Recent theoretical results were found for phases in which there are no
glass like properties. However, one may expect their validity in a modified
form also for the orientational glasses of the $RADP$ and $DRADP$ type in
which clusters with modulated structure are observed as we discussed above.
The presence of a frozen-in characteristic modulation wave vector \( q_{m} \)
at a given concentration x gives lower boundary \( ( q>q_{m} ) \) for those
wavelengths for which the response of the $RADP$ ($DRADP$) compounds may be
expected to be similar to that of the incommensurate phase. On the other
hand one may expect that response of long wavelength excitations of the
glass state gives information about an almost homogeneous effective medium with
properties averaged over all clusters. The transition region between
these two characteristic lengths behaves depending on interactions between
clusters. Note that the external probing perturbation characterized by
long wavelengths induces response of internal modes with very short wavelengths
due to coupling between long wavelength and short wavelength modes in the
incommensurate phase. One may then expect that mentioned above glass
response may be partially the same as that of the incommensurate phase
in corresponding ( averaged over impurities ) effective medium.

While the effective medium theory enables discussion of physical
properties on the macroscopic level, considerations
how the ground state modifies due to presence of defects on
the microscopic level are necessary to perform to understand
glassy state more thoroughly.

It is natural to extend
the studies of the sinusoidaly modulated
incommensurate structure in quantum structural order-disorder type materials
as idealized crystals, as in \cite {13}, to more realistic crystals with
a few deffects. Results of such studies may serve to be a starting point
to describe materials with nonzero concentration of defects and to
understand better limitations of the effective medium approach to such
systems.

The ideal systems without defects are
described by the quantum Ising-like Hamiltonian in the simplest cases.

\section{Model.}

Our description of the quantum order-disorder type of phase transitions
is based on the pseudospin formalism. The basic part,
\( H_{1} \), of our model Hamiltonian, \( H_{T} \), described
in \cite{14} for a lattice without defects, is given by
\begin{equation}
\label{1}
H_{1} = - \frac{1}{2} \sum_{i,j} J_{ij} S^{z}_{i} S^{z}_{j}.
\end{equation}

Here \( S^{z}_{i} \) are quantum operators describing
pseudospin variables. Their eigenvalues are
\( S^{z}_{i} = \pm \frac{1}{2} \), at lattice sites i, where
\( i = 1,...,N_{latt} \).
For simplicity
we study a s.c. lattice with \( N_{latt} \) sites,
in which each site represents a quantum
fluctuating unit (for example an hydrogen or deuteron bond).
The tunneling energy \cite{14} is
neglected in our model.

Real materials (for example of the $KDP$ type) are characterized by more
complicated structures and are described by
more complicated models, \cite{4}. We expect that our study
of the simplified model may give
results which will qualitatively demonstrate effects expected
to occur in real materials.

To formulate our model Hamiltonian we note that in the glassy materials
of the $RADP$ and $DRADP$ type one should consider a total Hamiltonian, \( H_{T} \),
which contains not only the basic part given above
but also the part \( H_{R} \) describing random fields \( h_{n} \) acting
on pseudospin degrees of freedom at those sites n at which substituents
\( ND_{4} \) are localized, \cite{4},
\begin{equation}
\label{2}
H_{R} = - \sum_{n} h_{n} S^{z}_{n}.
\end{equation}

In real systems it is necessary to
assume that resulting total field vanishes in the sample
\[       \sum_{n} h_{n} = 0.  \]

However for study purposes it will be instructive to understand
how presence of any field influences the ground state. In the next
sections we consider the case of a single impurity and a pair of
impurities characterized by a nonzero field.

In \cite{4}
the average lattice approximation is used as a first step in study of these
materials. It is
assumed that there exists a contribution to \( J_{ij} \) proportional to the
concentration of \( ND_{4}^{+} \) ions. In the present paper we will not
specify an explicit form of such contribution. It is however important that
we assume presence of
a competition between the ferroelectric (for x=0) and antiferroelectric
(x=1) interactions leading to frustration and resulting in modulation in
the pure lattice (without defects) ground state.

We assume
that the modulation occurs in the direction $(1,0,0)$ only.
The modulation wavevector amplitude \( Q \equiv {\bf Q.a} \), where
\( {\bf a} \) is the basic lattice vector in the direction $(1,0,0)$,
is determined by the free energy minimization. The interaction energy
Fourier transform $J(q)$ is assumed to has the form:
\[      J({\bf q}) = J_{1} \cos({\bf qa}) + J_{2} \cos(2 {\bf qa}). \]

We neglect its dependence on wavevectors in directions perpendicular
to the direction \( {\bf a} \). Such a step is valid
assuming weak modulations in these directions. We take
the interaction energies in perpendicular directions
to be zero.
The interaction energy constants \( J_{1} \)
and \( J_{2} \)
describe interaction between the nearest and next nearest neighbors
respectively.

\section{Ground state in pure systems.}

The random-phase-approximation type analysis, \cite{14}, of the basic
part (\ref{1}) of the total Hamiltonian gives description of the
modulated state. We perform in this section
such an analysis for reference purposes.
The free energy is minimized at a given temperature T for those values of the
pseudospin operators which satisfy
\begin{equation}
\label{3}
<S^{x}_{n}>=<S^{y}_{n}>=0,
\end{equation}
\[ <S^{z}_{n}>= \frac{1}{2} \tanh( \beta \frac{H_{n}}{2}), \]
\[ H_{n} = \sum_{j} J_{nj}<S^{z}_{j}> + h_{n}. \]

Note, that a random field influence due to defects is taken into account
in the internal field definition in (\ref{3}). Let us now consider thermodynamic properties of our model system for pure systems without defects.
The ground state equations (\ref{3}) give zero mean values of the operators
above some critical temperature \( T_{c} \) for zero random fields,
\( h_{n} = 0. \)
Below this temperature, which is given by
\begin{equation}
\label{4}
kT_{c} \equiv \frac{J(Q)}{2},
\end{equation}

the single plane wave modulated state
\begin{equation}
\label{5}
<S^{x}_{n}>=<S^{y}_{n}>=0,
\end{equation}
\[ <S^{z}_{n}>=  \cos( Qn + \phi ), \]

realizes in pure systems without random fields.
The modulation vector $Q$ is found in this case to be given by
\begin{equation}
\label{6}
\cos(Q) \equiv -\frac{J_{1}}{4J_{2}}.
\end{equation}

The internal mean field at the site n, \( H_{n} \),
 from (\ref{3}) is found to be given by
\[  H_{n} = 2SJ(Q) \cos(Qn + \phi ).  \]

We see that the amplitude of the internal field is proportional
to the amplitude of the pseudospin variable at a given site in the
single plane wave state. This later property is lost when higher harmonics
terms are included in the modulated state (\ref{5}).

The wavevector $Q$ is incommensurate if $Q$ is different
from \( 2 \pi \frac{M}{N} \), here M, N are any integers.
Note that the case of $DRADP$ corresponds to \( \frac{Q}{2 \pi} \approx 0.35 \).
The case of \( Rb_{2}ZnBr_{4} \) may be characterized by \( \frac{Q}{2 \pi}
\approx \frac{5}{17} \), see in \cite{17}.

The ground state energy is given by
\begin{equation}
\label{7}
E_{GS} = - N_{latt} J(Q) \frac{S^{2}}{2},
\end{equation}

where the amplitude of the wave, S, is found from (\ref{3}) - (\ref{5})
\begin{equation}
\label{8}
S^{2} \equiv \frac{T^{3}}{T^{3}_{c}}( \frac{T_{c}}{T} - 1).
\end{equation}

The validity of the expression (\ref{8}) is restricted to
temperatures not very low with respect to the transition temperature,
the region where the single plane wave dominates. One should modify (\ref{8})
whenever higher harmonics become essential. In some systems the
basic harmonics may remain the most important harmonics also
at low temperatures. Glassy materials in which freezing
temperature is nearby the transition temperature given above are
one example. Another example are systems in which some of higher harmonics
are suppressed due to symmetry even at low temperatures.
In systems described by the Hamiltonian given above and in which
there are no reasons to suppress higher harmonics
the third harmonic amplitude becomes very soon comparable to the
basic harmonic one. Its amplitude \( S_{3} \) is given by
\[ S_{3} \equiv S^{3} . \frac{\frac{-T_{c}^{3}}{3T^{3}} }{1-
\frac{J(3Q)}{2k_{B}T}} . \]

Here and in the following we assume that the modulation wavevector
amplitude $Q$ leads to the incommensurate modulated phase. Thus we will not
discuss the effects of the lock-in energies.

\section{Single plane wave ground state and impurities.}

The Hamiltonian part \( H_{R} \) describing random fields \( h_{n} \) acting
on pseudospin degrees of freedom at sites n (at which substituents
\( ND_{4} \) are localized) has form (\ref{2}), see \cite{4}. In
the mean field approximation the energy \( E_{R} \) associated with this part
of the total Hamiltonian has the form
\begin{equation}
\label{2.1}
E_{R} = - \sum_{n} h_{n} < S^{z}_{n} >.
\end{equation}

While in real systems it is necessary to
assume that resulting total field in the sample vanishes
\(       \sum_{n} h_{n} = 0  \), it is not the case of a single
defect case study.
We further assume that the modulated ground state has a single
plane wave form
\[ < S^{z}_{n} > = S.cos(QD+\phi).     \]

This assumption is correct when random fields have a very small amplitude.
As it will be discussed later, in this case the equations determining the
form of the ground state have the same form as those in pure material
in the highest of H order. In the same order the solution to these equations
describing the ground state is of the same order. Thus we can use
(\ref{2.1}) to study influence of defects on the ground state in the
highest in h order.
Our aim is to determine which values of the wave amplitude S,
the wave vector $Q$ and the phase variable \( \phi \) minimize
the free energy at given thermodynamic conditions.
The free energy is minimized (at a temperature T) for those values of the
pseudospin operators which satisfy
\begin{equation}
\label{3.1}
<S^{x}_{n}>=<S^{y}_{n}>=0,
\end{equation}
\[ <S^{z}_{n}>= \frac{1}{2} \tanh( \beta \frac{H_{n}}{2}), \]
\[ H_{n} = \sum_{j} J_{nj}<S^{z}_{j}> + h_{n}. \]

In this paper we would like to consider systems with one and two
defects. Then we may proceed our calculations as in the pure
systems taking \( h_{n}=0 \) in (\ref{3.1}). Influence of
defects may be done in such case by considering the energy terms
arising from interactions of pure system with defects.

The ground state equations (\ref{3.1}) give zero mean values of the operators
above the critical temperature \( T_{c} \). Below this temperature,
given by (\ref{4}), which has the same value as in the pure case,
the single plane wave modulated state
\begin{equation}
\label{5.1}
<S^{x}_{n}>=<S^{y}_{n}>=0,
\end{equation}
\[ <S^{z}_{n}>=  \cos( Qn + \phi ), \]

realizes again.
The modulation vector $Q$ is found to be again, as in the pure case, given by
\begin{equation}
\label{6.1}
\cos(Q) \equiv -\frac{J_{1}}{4J_{2}}.
\end{equation}

The internal mean field at the site n, \( H_{n} \),
from (\ref{3}) is found to be given by
\[  H_{n} = 2SJ(Q) \cos(Qn + \phi ) + h_{D} \delta(n,D),  \]

where D are sites where defects are localized.

\subsection{A single impurity.}

Let us assume that the lattice site D contains a defect and that
the field associated with this defect is given by \( h_{D} \equiv h. \)
The energy \( E_{R}^{s} \) associated with this part
of the total Hamiltonian has the form
\begin{equation}
\label{2.2}
E_{R}^{s} = -  h.S.cos(QD+\phi).
\end{equation}

Let us first consider influence of this energy term when the amplitude
of the field h is very small and positive. Turning on slowly this field
we see from (\ref{2.2}) that the phase variable \( \phi \) locks
its value even in the incommensurate phase to
\[ \phi = -QD + 2 \pi k, \]

where k is any integer.
The amplitude S remains the same as in the pure case in the first
order of the (small) field h.
From (\ref{2.2}) it is clear that such a defect will have tendency
to suppress the amplitude S to lower values.

\subsection{A pair of impurities.}

Let us assume that the lattice sites $0$ and $D$ contain a pair of defects.
The fields associated with these defects are given by \( h_{0} \equiv h, \)
\( h_{D} \equiv -h. \) We assume that the sum of the fields on defects
is zero.
The energy \( E_{R}^{s} \) associated with this part
of the total Hamiltonian has the form
\begin{equation}
\label{2.2.1}
E_{R}^{s} = +  h.S.cos(QD+\phi) - h.S.cos(\phi).
\end{equation}

This term is the only term in the total energy which is dependent
on the phase variable \( \phi. \) Thus its minimization with
respect to this phase variable gives description about its thermodynamic
state.

Let us consider influence of this energy term when the amplitude
of the field h is very small and positive. Turning on slowly this field
we see from (\ref{2.2.1}) that the phase variable \( \phi \) locks
its value even in the incommensurate phase to
\[ \phi = \frac{\pi}{2}- \frac{QD}{2} + 2 \pi k, \]

where k is any integer.
The amplitude S remains the same as in the pure case in the first
order of the (small) field h.
From (\ref{2.2.1}) it is clear that such a pair of defects will have tendency
to suppress the amplitude S to lower values:
\begin{equation}
\label{2.2.2}
E_{R}^{s} = - h.S.\vert \cos( \frac{QD}{2} ) \vert.
\end{equation}

This energy (\ref{2.2.2}) is always a negative quantity for h positive
a D nonzero, if $Q$ takes an incommensurate value.

\section{Third harmonics of the ground state and impurities.}

\subsection{A single impurity.}

When the third harmonics in the formation of the ground state
in pure system is taken into account, we obtain the ground
state
\[ <S^{z}_{n}>= S \cos( Qn + \phi ) + S_{3} \cos(3 Qn + 3 \phi ), \]

where temperature dependence of amplitudes S and \( S_{3} \) was
shown in the Introduction.

Let us assume that the lattice site D contains a defect and that
the field associated with this defect is given by \( h_{D} \equiv h. \)
The energy \( E_{R}^{s} \) associated with this part
of the total Hamiltonian has the form
\begin{equation}
\label{2.2.a}
E_{R}^{s} = -  h.S.cos(QD+\phi) -  h.S_{3}.cos(3QD+3 \phi).
\end{equation}

This energy is small with respect to the energy of the whole lattice.
This later energy, however, determines both amplitudes. Thus
we assume that the amplitudes and the wave vector $Q$ are unchanged with
respect to the pure case.
Let us first consider influence of this energy term when the amplitude
of the field h is very small and positive. Turning on slowly this field
we see from (\ref{2.2.a}) that the phase variable \( \phi \) locks
its value even in the incommensurate phase to
\[ \phi = -QD + 2 \pi k, \]

where k is any integer. This later statement is true for temperature
near the critical temperature, where
\[ \vert S \vert >> \vert S_{3} \vert.  \]

However, there may exist metastable (a local minimum of free energy) states
given by
\[ \phi = -QD + 2 \pi k + \phi_{0}, \]

where temperature dependence of this phase is given by
\[ \sin(\phi_{0})^{2} =  \frac{3}{4} - \frac{S}{12 \vert S_{3} \vert }. \]

These states exist below some temperature lower than the critical.
We will not discuss them further in this paper.

The amplitude S remains the same as in the pure case in the first
order of the (small and positive) field h. It is clear that
while the basic harmonics amplitude should increase when
the single impurity is present, the third harmonics amplitude
will have tendency to decrease its value
\begin{equation}
\label{2.2.2.a}
E_{R}^{s} = -  h.S -  h.S_{3},
\end{equation}

due to fact that the sign of both amplitudes is reversed.

\subsection{A pair of impurities.}

Let us assume that the lattice sites 0 and D contain a pair of defects.
The fields associated with these defects are given by \( h_{0} \equiv h \)
and \( h_{D} \equiv -h. \)
The energy \( E_{R}^{s} \) associated with this part
of the total Hamiltonian has the form
\begin{equation}
\label{2.2.3}
E_{R}^{s} = +  h.S.cos(QD+\phi) +  h.S_{3}.cos(3QD+3 \phi)-
\end{equation}
\[ -  h.S.cos( \phi) -  h.S_{3}.cos(3 \phi). \]

Here we assume that the ground state is described by the same phase
variable in the whole crystal. This energy is again small with respect to the energy of the whole lattice.
We assume that the amplitudes and the wave vector $Q$ are unchanged with
respect to the pure case.
Let us first consider influence of this energy term when the amplitude
of the field h is very small and positive. Turning on slowly this field
we see from (\ref{2.2}) that the phase variable \( \phi \) locks
its value even in the incommensurate phase to
\[ \phi = -\frac{QD}{2} + \frac{ \pi}{2}, \]

where k is any integer. This later statement is true for temperature
near the critical temperature, where
\[ \vert S \vert >> \vert S_{3} \vert  \]

and where
the energy of the ground state modulated by basic and the third
harmonics and originated from interaction with defects,
found to be given by
\begin{equation}
\label{2.2.2.b}
E_{R}^{s} = -  2.h.S \sin(\frac{QD}{2})+  h.S_{3} \sin(\frac{3QD}{2}),
\end{equation}

is negative.
We see that this happens for those distances between two defets for
which
\[ \sin(\frac{QD}{2}) >0,\]
\[ \sin(\frac{3QD}{2})<0,\]

due to fact that the sign of both amplitudes is reversed.
Also the case
\[ \sin(\frac{QD}{2}) <0,\]
\[ \sin(\frac{3QD}{2})>0,\]

gives the same energy, the phase variable is now shifted by \( \pi \).
There is however region of ditances between defects for which
the second term in (\ref{2.2.2.b}) is positive. In case of two defects
this does not suppress the third harmonics.

\section{Discussion.}

The energy \( E_{R}^{s} \) associated with interaction of two defects
with modulated ground state wave consisting of the basic and the third
harmonics has the form given by (\ref{2.2.2.b}). Let us consider a medium
in whic such defects are distributed more-less homogeneously
in all directions. If x is their density and D is their average distance
in the modulation direction, then the total excess energy due to interaction
of considered defects with the ground state modulated by the basic and
the third harmonics is found to be given by
\begin{equation}
\label{2.2.4.A}
E_{R}^{s} = -  2.h.S.N.x. \sin(\frac{QD}{2})+  h.S_{3}.N.x \sin(\frac{3QD}{2}) -
N.S_{3}^{2} \frac{J(3Q)}{2},
\end{equation}

where the last term is due to the third harmonics contribution to the total
internal energy.
Let us compare this energy
with the energy of the same state in which the third harmonics is absent,
\( S_{3}=0\).
This later energy excess will be
\begin{equation}
\label{2.2.3.A}
E_{R}^{s} = -  2.h.S.N. \vert \sin( \frac{QD}{2}) \vert .
\end{equation}

Here we assume that the ground state is described by the same basic
harmonics in the whole crystal.

There exists relationship between the defects average distance in the
modulation direction D, the defects average distance in the perpendicular to the
modulation direction \( D_{p} \), the lattice constan a and the defects
concentration x:
\[ x= \frac{a^{3}}{D.D_{p}^{2}}. \]

The third harmonics becomes suppressed  when the difference
of these two energies (\ref{2.2.4.A}) and (\ref{2.2.3.A})
becomes comparable.

We see that this happens for those distances between two defets for
which either basic harmonics
(which still dominates the third harmonics)
in the presence of defects and the third harmonics contributes
a large amount of energy when
\[ \sin(\frac{QD}{2}) >0\]

either when
\[ \sin(\frac{QD}{2}) <0\]

and simultaneously
\[   h.S_{3}.x \sin(\frac{3QD}{2}) -
S_{3}^{2} \frac{J(3Q)}{2} >0. \]

When the inequalities just displayed are not satisfied for both cases
the third harmonics is present in the system.

We hope that our theoretical
discussion of possibility to find the above mentioned
type of modifications of the ground state structure due to disturbancies
in materials of the quantum order-disorder
type ($KDP$-like materials) in which incommensurably modulated phase may occur
enable to increase the number of types of materials in which
studied phenomena may be confronted with reality.


\begin{thebibliography}{MLM9211}
\bibitem{tfc}T.F. Connolly, Ferroelectrics Literature Index, ISBN:
1-46846-210-5, Publisher: Springer Science and Business Media, 
p. 705 (2012)
\bibitem{WBJW}W. Benenson, J.W. Harris, H. Stöcker and H. Lutz, Handbook of Physics, ISBN: 0-387-95269-1, Publisher:Springer Science and Business Media, p. 1186, (2006)
\bibitem{AMEC}A. Milsted, E. Cobanera, M. Burrello and G. Ortiz, Phys.Rev. {\bf B90} p. 195101 (2004)
\bibitem{YLMAG}Y. Lansac, M.A. Glaser and N.A. Clark, Phys.Rev. {\bf E73} p. 041501 (2006)
\bibitem{GDGHR}G.D. Granzow and H. Riecke, Physica A: Statistical Mechanics and its Applications {\bf 249} Ns. 1–4 pp. 27-35 (1998)
\bibitem{ANRTJ}A.N. Rubtsov and T. Janssen, Thermodynamic Properties of the Discommensuration Point for Incommensurate Structures: A "Third-Order" Phase Transition, arXiv ID: cond-mat/0304447 (2003)
\bibitem{SAMM}S. Albert, M. Michl, P. Lunkenheimer, A. Loidl, P. Déjardin and F. Ladieu, Third and Fifth Harmonic Responses in Viscous Liquids, In: Richert R. (eds) Nonlinear Dielectric Spectroscopy. Advances in Dielectrics. Springer, Cham (2018)
\bibitem{JLYN}J. Lin, Y. Ni, Zh. Hao, J. Zhang, W. Mao, M. Wang, W. Chu, R. Wu; Zh. Fang, L. Qiao, W. Fang, F. Bo and Y. Cheng, Highly-efficient second and third harmonic generation in a monocrystalline lithium niobate microresonator, arXiv ID: 1809.04523 (2018)
\bibitem{AYBPKD}A.Y. Borisevich and P.K. Davies, Journal of American Ceramic Society {\bf 85} N. 3 pp. 573-578 (2002)
\bibitem{31}B.A.Strukov, Phase Transitions {\bf 15} pp. 143-179 (1989) 
\bibitem{32}J.P.Jamet, in Competing Interactions and Microstructures:
Statics and Dynamics, R. LeSar, A.Bishop and R.Heffner, Springer Proc.
in Physics {\bf 17} (1988)
\bibitem{33}T.Natterman, Ferroelectrics {\bf 104} pp. 171-181 (1990) 
\bibitem{5}H.Grimm and J.Martinez, Z.Phys. B - Condensed Matter {\bf 64} p. 13
(1986) 
\bibitem{3}R.A.Cowley, T.Ryan and E.Courtens, J.Phys.C: Solid State Physics
{\bf 18} p. 2793 (1985) 
\bibitem{10}J. Petzelt, V. Zelezny, S.Kamba, A.V.Sinitski, S.P.Lebedev,
A.A.Volkov, G.V.Kozlov and V.H.Schmidt, J.Phys.: Condensed Matter {\bf 3} p. 2021 (1991)
\bibitem{MLM 92}M.LeMaire, G.Lingg, G.Schaack, M.Schmitt-Lewen and
G.Strauss, Ferroelectrics {\bf 125} p. 87 (1992) 
\bibitem{ctdlh}C. Thibierge, D. L'Hote, F. Ladieu and R. Tourbot, Rev.Sci.Instrum. {\bf 79} N. 10 p. 103905 (2008)
\bibitem{14}R.Blinc and B.Zeks, Soft Modes in Ferroelectrics and
Antiferroelectrics, North-Holand Pub.Comp., Amsterdam, Oxford, ch.V. (1974)
\bibitem{15}S.W.Lovesey, J.Phys.C : Solid State Phys. {\bf 21} p. 2805 (1988)
\bibitem{13}S.W.Lovesey, G.I.Watson and D.R.Westhead, Int.J.Mod.Physics
{\bf 39} p. 405 (1990) 
\bibitem{DHH}O. Hudak, V. Dvorak, J. Holakovsky and J. Peztelt,
Phys. Scr. {\bf 55} p. 77 (1994)
\bibitem{1}U.T.Hoechli, K.Knorr and A.Loide, Advances in Physics,
{\bf 39} p. 405 (1990) 
\bibitem{2}E.Courtens, T.F.Rosenbaum, S.E.Nagler and P.M.Horn, Phys.Rev. {\bf B29} p. 515 (1984) 
\bibitem{4}R.A.Cowley, T.W.Ryan and E.Courtens, Z.Phys. B - Condensed
Matter {\bf 65} 181 (1986)
\bibitem{6}H.Grimm, K.Parlinski, W.Schweika, E.Courtens and H.Arend,
Phys.Rev. {\bf B33} p. 4969 (1986)
\bibitem{7}H.Grimm, in Dynamics of Disordered Materials, Springer Proceedings
in Physics {\bf 37} p. 274, Eds. D.Richter, A.J.Dianonx, W.Petry and J.Teixeira,
Springer-Verlag Berlin-Heidelberg, (1989) 
\bibitem{8}P.He, K.Deguchi, H.Hirokane and E.Nakamura, J.Phys.Soc.Japan
{\bf 59} p. 1835 (1990) 
\bibitem{9}J.-F. Berret, M. Meissner, S.K. Watson, R.O. Pohl and E. Courtens,
Phys.Rev.Lett. {\bf 67} p. 93 (1991) 
\bibitem{11}S.Kamba, J.Petzelt, A.V.Sinitski, A.G.Pimenov, A.A.Volkov and G.V.Kozlov,
Ferroelectrics {\bf 127} N. 1 pp. 263-268 (1992) 
\bibitem{12}P. He, J.Phys.Soc.Japan {\bf 60} p. 313 (1991)
\bibitem{17}H.Z.Cummins, Phys.Reps. {\bf 185} p. 321 (1990)  
\end{thebibliography}
\end{document}